\documentclass[a4paper,10pt]{article}

\usepackage{graphicx}
\usepackage{booktabs,textcomp}

\usepackage{geometry}
\usepackage{amsmath,cite}
\usepackage{amssymb,tocloft}
\usepackage{hyperref}

\begin{document}
\title{Reionization effect enhancement due to primordial black holes}
	
\author{K.~M.~Belotsky$^1$, A.~A.~Kirillov$^{1,2}$, N.~O.~Nazarova$^1$\thanks{nazarova.mephi@gmail.com}, S.~G.~Rubin$^{1,3}$\\
{$^1$National Research Nuclear University ``MEPhI'' 
 (Moscow Engineering Physics Institute),}\\
Kashirskoe shosse 31, Moscow, 115409, Russia.\\
$^2$ Yaroslavl State P.~G.~Demidov University,\\
  Sovietskaya street 14, Yaroslavl, 150003, Russia.\\
$^3$ N.~I.~Lobachevsky Institute of Mathematics and Mechanics,\\
Kazan  Federal  University, \\
Kremlevskaya  street  18,  420008  Kazan,  Russia}

\date{}

\maketitle
		
\begin{abstract}
	Primordial black holes (PBH) could account for variety of phenomena like dark matter, reionization of the Universe, early quasars, coalescence of black holes registered through gravitational waves recently. Each phenomenon relates to PBH of a specific mass range. PBH mass spectra varies in a wide range depending on specific model.  Earlier we have shown that PBH with monochromatic mass distribution around $5\times 10^{16}$ g value allow to re-ionize the Universe moderately. 
	Here we show that reionization effect and contribution to dark matter can be simultaneously enhanced with more natural extended mass distribution in the range around the same mass value.
\end{abstract}

\section*{Introduction}

There are many mechanisms of primordial black hole (PBH) formation. Specific feature of such objects is that they 
can be formed in a very broad mass range even if a mechanism is chosen. Depending on mass, PBHs could play different roles in cosmology and astrophysics. PBHs with mass ranging $1-1000 M_\odot$ could account for both (appreciable part of) dark matter (DM) and gravitation event GW150914 \cite{GW_observ,GW}, while contribution in DM becomes more constrained \cite{abs, Carr, 0912.5297}. Mechanisms described in \cite{1, Dolgov, Carr, 0912.5297} may lead to formation of supermassive black holes. In the works~\cite{3,32,33}, we explored the Hawking radiation of PBH for an explanation of the reionization at redshift $z\sim 8$, which proved by the different observations \cite{Planck1, Planck2}. We found that for a delta-function-like mass distribution, the effect can be reached within narrow mass interval around $5\times 10^{16}$ g but 
only for 
$z\lesssim 4$. 
This mass interval is close to that where PBHs can contribute noticeably to the density of DM, while explanation of all DM could require specific adjusting of PBH mass spectrum in this range \cite{Carr, 0912.5297}. 



There is a set of mechanisms (see reviews \cite{Carr, 0912.5297,33,Khlopov}) leading to a variety of PBH mass spectra. In this paper we consider those spectra of PBH masses that could explain the early reionization without connection to specific model.
We study and compare contributions to the reionization and dark matter of the Universe for different mass distributions like delta-functions, power law distributions (including falling, growing and uniform). The appropriate PBH mass interval is  $10^{15} \, \text{g}\lesssim M \lesssim 10^{18}~$g. Note that these PBHs could also explain positron line from Galactic center \cite{Gamma-line_1} due to effects of accretion \cite{Gamma-line_2} or Hawking evaporation \cite{Gamma-line_3}.
				
Constraints on the PBH density are usually applied only for delta-function mass distributions \cite{Carr, 0912.5297}. 
In the mass range of our current interest, constraint comes mainly from the observed diffuse gamma-ray background~(DGRB)~\cite{Carr, 0912.5297}. We reproduced it in figure~\ref{constrain} (left). The density value is given in term of ratio of cosmological PBH density $\Omega_{\rm PBH}$ to $\Omega_{\rm CDM}\approx 0.26$.
At high mass tail, $M\gtrsim 10^{17}$~g, constraint from so called femtolensing \cite{femt} starts to prevail. We do not show it here\footnote{The lower edge of mass interval where femtolensing constraint comes into force is indicated as from $5\times 10^{16}$~g to $5\times 10^{17}$~g in different articles.} and put for the most of calculations upper limit for PBH mass distribution to be $M_{\max}=M_{17}=10^{17}$ g, to weaken impact of femtolensing constraint (FC) or evade it. But special cases when $M_{\max}\lesssim M_{17}$ and $M_{\max}> M_{17}$ will be discussed.

For extended mass distribution of PBHs, we get upper limit on the PBH density by comparing the estimated Hawking gamma radiation of PBHs in the full mass range with the data from HEAO, COMPTEL and EGRET~\cite{20,5}, keeping in mind that $\Omega_{\rm PBH}\le \Omega_{\rm CDM}$\footnote{Spectrum of gamma radiation from single PBH is approximated by the Planck black body formula multiplied by a polynomial in energy to fit the expected flux and reproduce constraint of \cite{Carr, 0912.5297} in the given range as shown in figure~\ref{constrain} (left).}.
Then we evaluate the contribution of Hawking radiation to the reionization of the Universe along with contribution to DM density, which is provided by PBHs from all mass interval. 
				
PBHs at masses of interest have Hawking temperature within the interval $10 \,\text{keV} \lesssim T_{\text{PBH}} \lesssim 10 \, \text{MeV}$ so emit \cite{23} gamma rays $\gamma$, electrons and positrons $e^{\pm}$, neutrinos $\nu_{e,\mu,\tau}$ and gravitons $G$. 		
Ionization losses of $e^{\pm}$ provide the main contribution to the ionization effect of matter, which can be assumed to proceed homogeneously in space \cite{3}. 
Gamma rays from PBHs are not so effective, they mostly provide imposing limits on the PBH density by their contribution to the DGRB\footnote{Nevertheless the gamma-radiation can provide the observational effects explaining unidentified point-like gamma-ray sources within Galaxy \cite{PBH_PGRS, PBH_PGRS_2}.}.
		
 \section*{Basic formulas}
 
In calculation of the temperature $T$ and  ionization degree $x_e$ of baryonic matter, we follow here the work \cite{3}. Note that the approximation used there is not so quantitatively correct (however basically when $x_e\ll 1$ what is rather not of interest) \cite{30,31}, but nonetheless seems to be, at least, qualitatively acceptable \cite{31} for estimation of reionization effect and, what is our aim here, demonstration how the effect can be \textit{relatively} enhanced due to extended PBH mass spectra.

The main difference of current calculations from those of \cite{3} is that we use distribution of PBHs in mass. To take into account this one takes Eq.(5) of \cite{3} and generalizes it for a case of extended mass distribution:
\begin{equation}
\frac{d\dot{\Omega}_{\rm ev}}{dM}=\frac{\dot{M}}{M}\frac{d\Omega_{\rm PBH}(M)}{dM}
= \frac{1}{3}\left(\frac{M_U}{M}\right)^3\frac{d\Omega_{\rm PBH}(M)/dM}{t_U}.
\label{Omegaev}
\end{equation}
The sense of presented value is the energy evaporation rate per unit volume, divided by critical density ($\rho_{\rm crit}$). The value $\dot{M}$ is the energy evaporation rate of single PBH, $M_U\approx 5\times 10^{14}$ g is the mass of PBH which is evaporated completely for the modern age of Universe $t_U$.
Distribution of PBH cosmological density in $M$ can be connected with conventional probability distribution, normalized on unit, ($\frac{dw}{dM}$) as
\begin{equation}
\frac{d\Omega_{\rm PBH}(M)}{dM}=\frac{M}{\bar{M}}\Omega_{\rm PBH}\frac{dw}{dM},
\end{equation} 
where $\bar M\equiv \int_{M_{\min}}^{M_{\max}}M\frac{dw}{dM}dM$ is the mean mass, $\Omega_{\rm PBH}$ is the total density (of PBHs of all masses) to be found from DGRB and CDM density constraints.
Then we put the value \eqref{Omegaev} in 
Eq.(21) of \cite{3}, generalizing it analogously to \eqref{Omegaev}, 
which in its turn put to Eq.(27) through replacement there 
$$
	\dot\Omega_{\rm abs}\rightarrow \int_{M_{\min}}^{M_{\max}}\frac{d\dot\Omega_{\rm abs}^{(e-\text{ion})}}{dM}dM.
$$
Given value has the sense of absorbed energy by baryonic matter due to ionization process (other processes of energy transfer from evaporation products to the baryonic matter can be neglected \cite{3,33}). Note that it takes into account that energy absorption process takes a finite time so electrons emitted by the PBH have time to lose energy also due to scattering on CMB and due to the red shift.	
Finally, solution of Eq.(27) gives us the temperature, and Eq.(28) of \cite{3}~--- the ionization degree. 

The $\gamma$-ray flux from PBHs is estimated as 
\begin{gather}
F_{\gamma}^{\text{mod}}(E) = \frac{c}{4\pi} \, \rho_{\rm crit}  \iint \frac{ \kappa_{\gamma}}{\left\langle E_{\gamma} \right\rangle   } \, \frac{\dot{M}}{M} \, f_{\text{Pl}}(M,E_{\gamma0}=E_{\gamma}(z+1)) \frac{d\Omega_{\text{PBH}}}{dM} \, dM\, \frac{H^{-1}_{\rm mod}dz}{\sqrt{\Omega_{m} \left(z+1 \right)^{3} + \Omega_{\Lambda}}}.
\end{gather}
Here
$\kappa_{\gamma}$ is the energy fraction evaporated in form of $\gamma$, $E_{\gamma0,\gamma}$ are their initial (as radiated by PBH) and final (at the Earth) energy,
$\left\langle E_{\gamma} \right\rangle $ is their mean (final) energy,
$f_{\text{Pl}}$ is the initial photon spectrum, normalized on unit (modified Planck form),
$\Omega_{\Lambda}=0.69$ and $\Omega_m=0.31$ are the modern energy and non-relativistic matter densities,
$H_{\rm mod}$ is the modern Hubble parameter.
	
We do not consider here contribution in DGRB from PBHs in Galaxy. It leads to constraining abundance of basically less massive PBHs while suffers by extra uncertainties related with PBH distribution in Galaxy  \cite{1604.05349, 0912.5297}. In fact, PBHs in Galaxy may give relatively big contribution in gamma when they are already at the last active evaporation stage. It requires their initial mass to be tuned around $M_U$, what is well below of that needed here.
		
\section*{Mass spectra with two peaks}		
		
We start calculation with the single delta-function mass distribution to compare its result with those for other distributions.
		
\begin{figure}
   		\begin{center}  	
    			{\includegraphics[scale=0.3]{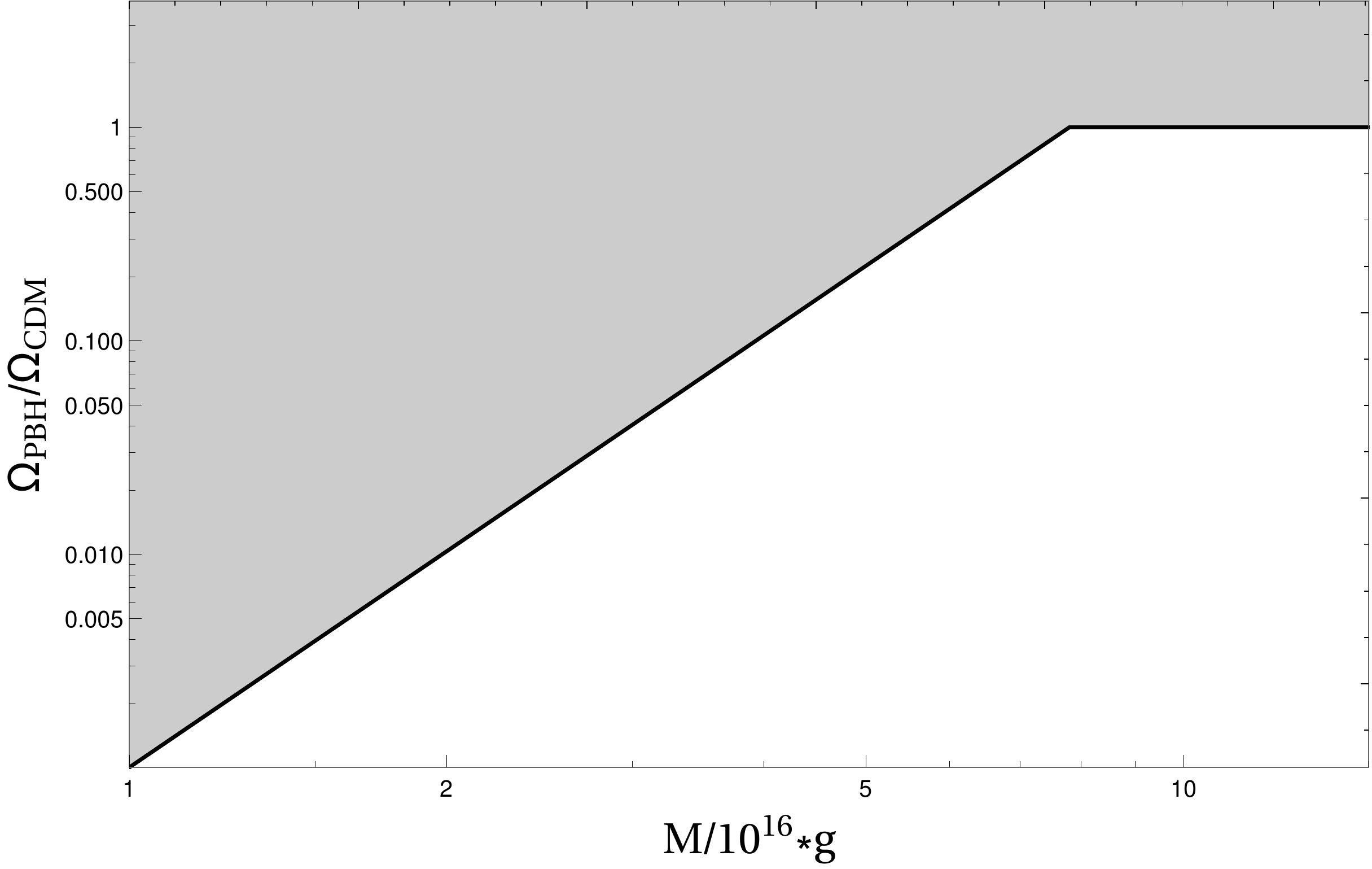}}
   			\quad
   			{\includegraphics[scale=0.3]{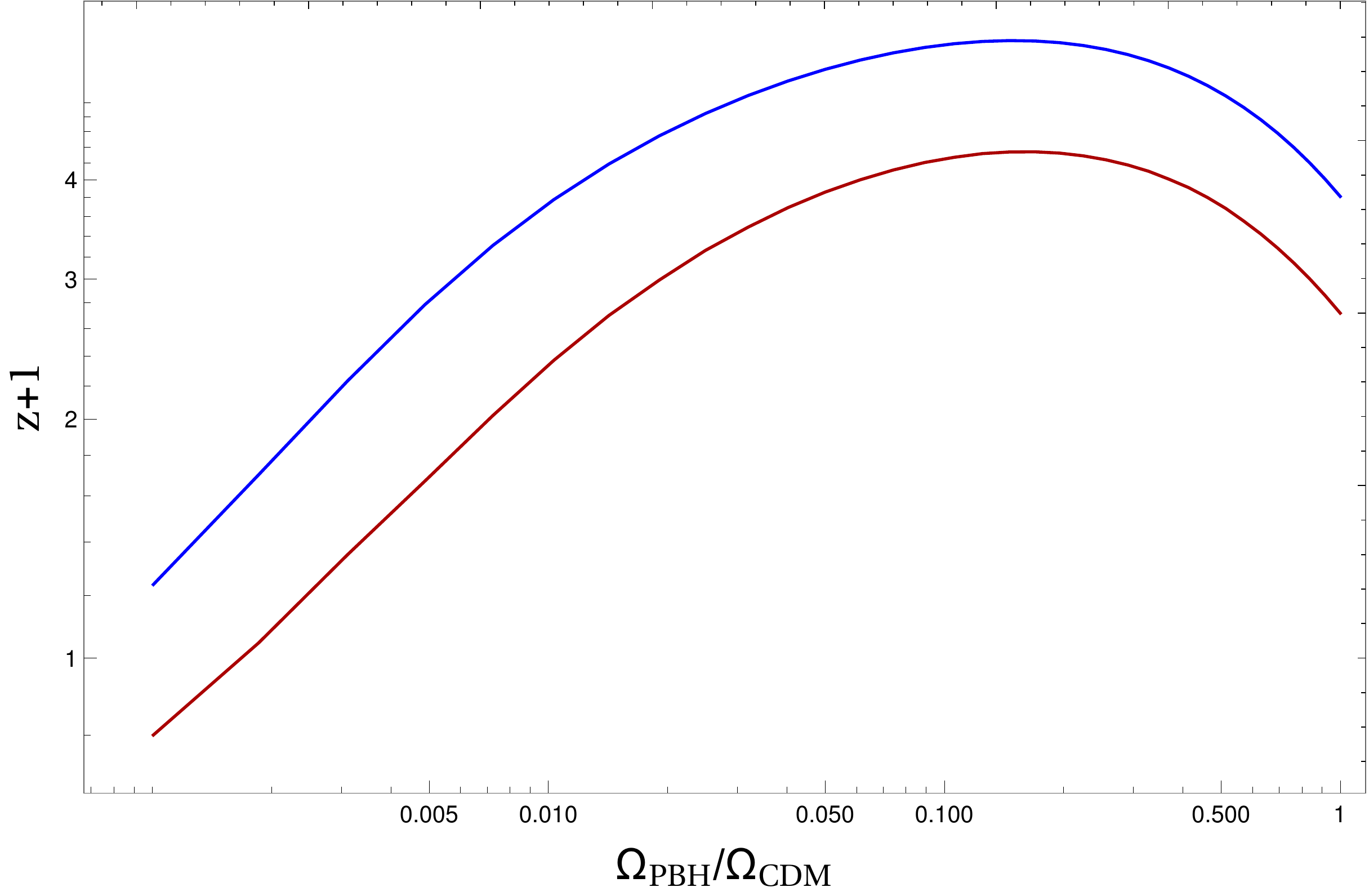}}
   		\caption{ (left) Constraints on the PBH density for the mass range of interest, obtained for the delta-function-like mass spectra. Within given mass range, constraint from DGRB prevails. Shaded region is forbidden.  
   		(right) Dependence of the redshifts, at which the ionization of matter achieves  80\% (blue line) and 100\% (red line), from PBH density, which they have accordingly to the upper limit shown left.} 
   		\label{constrain}
   			\end{center}
   	\end{figure}

Constraint on $\Omega_{\rm PBH}$ obtained from the data on gamma radiation, as mentioned above, is shown in figure~\ref{constrain} (left).
The figure~\ref{constrain} (right) shows the redshift, at which PBHs ionize  80\% (blue line) and 100\% (red line) of matter, depending on their contribution to dark matter as constraint (shown left) allows.
The x-axis corresponds to the PBH mass range   $10^{16} \, \text{g}\lesssim M \lesssim 8 \times 10^{16} \,$g (in this range, maximal $\Omega_{\rm PBH}$ is uniquely defined by $M$).
As clearly seen, the solution of the problems of dark matter and reionization of the Universe can not be reached simultaneously with the single delta-function PBH mass distribution.
For largest contribution to reionization ($z\sim 4$), not more than 40--50\% of dark matter can be in the form of PBHs.
On the contrary, if all dark matter density consists of PBHs, the reionization happens (due to PBH Hawking radiation) not earlier than $z\sim 3$.

\begin{figure}
		\begin{center}
			\includegraphics[scale=0.3]{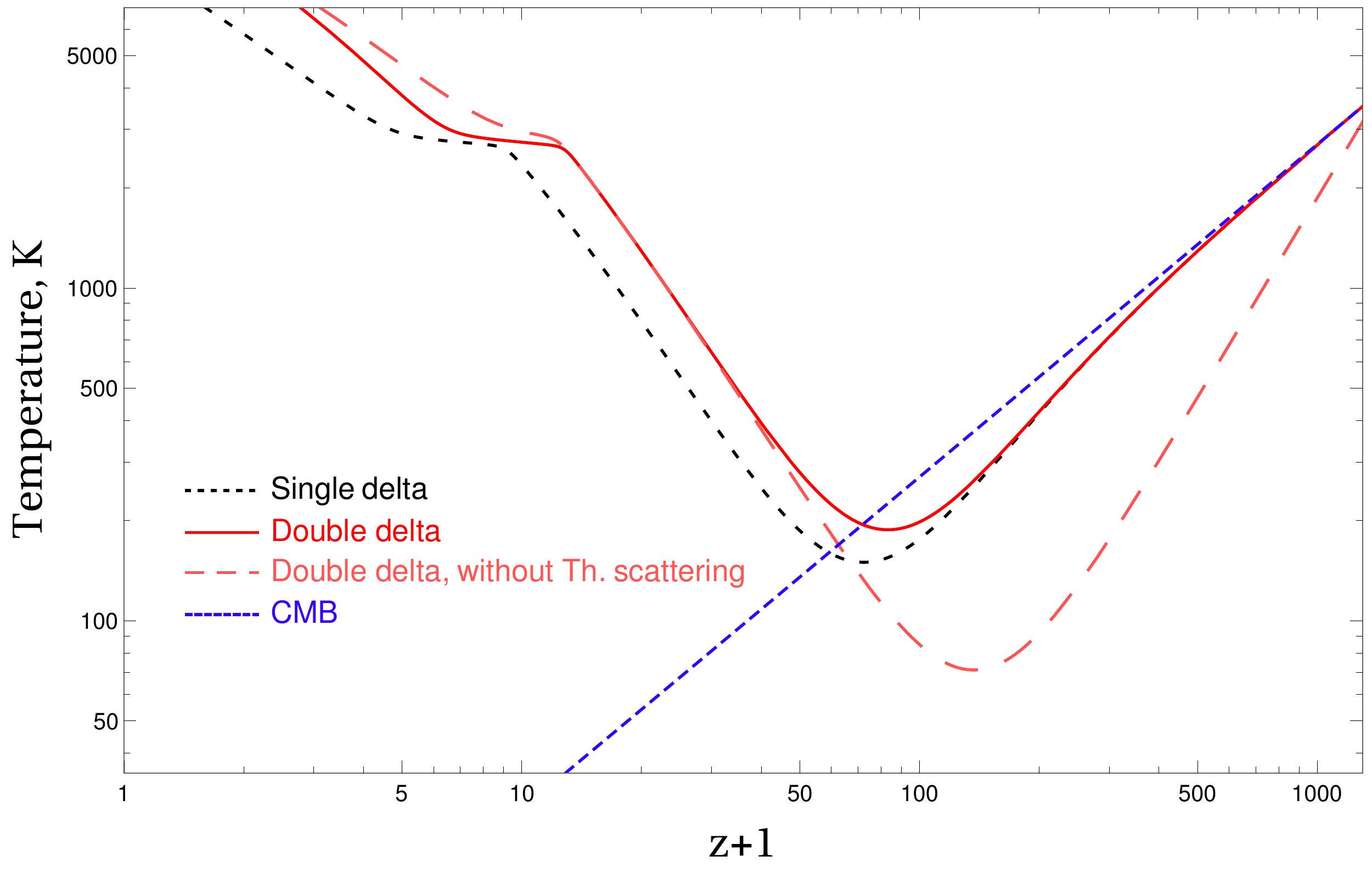}
			\quad
			\includegraphics[scale=0.3]{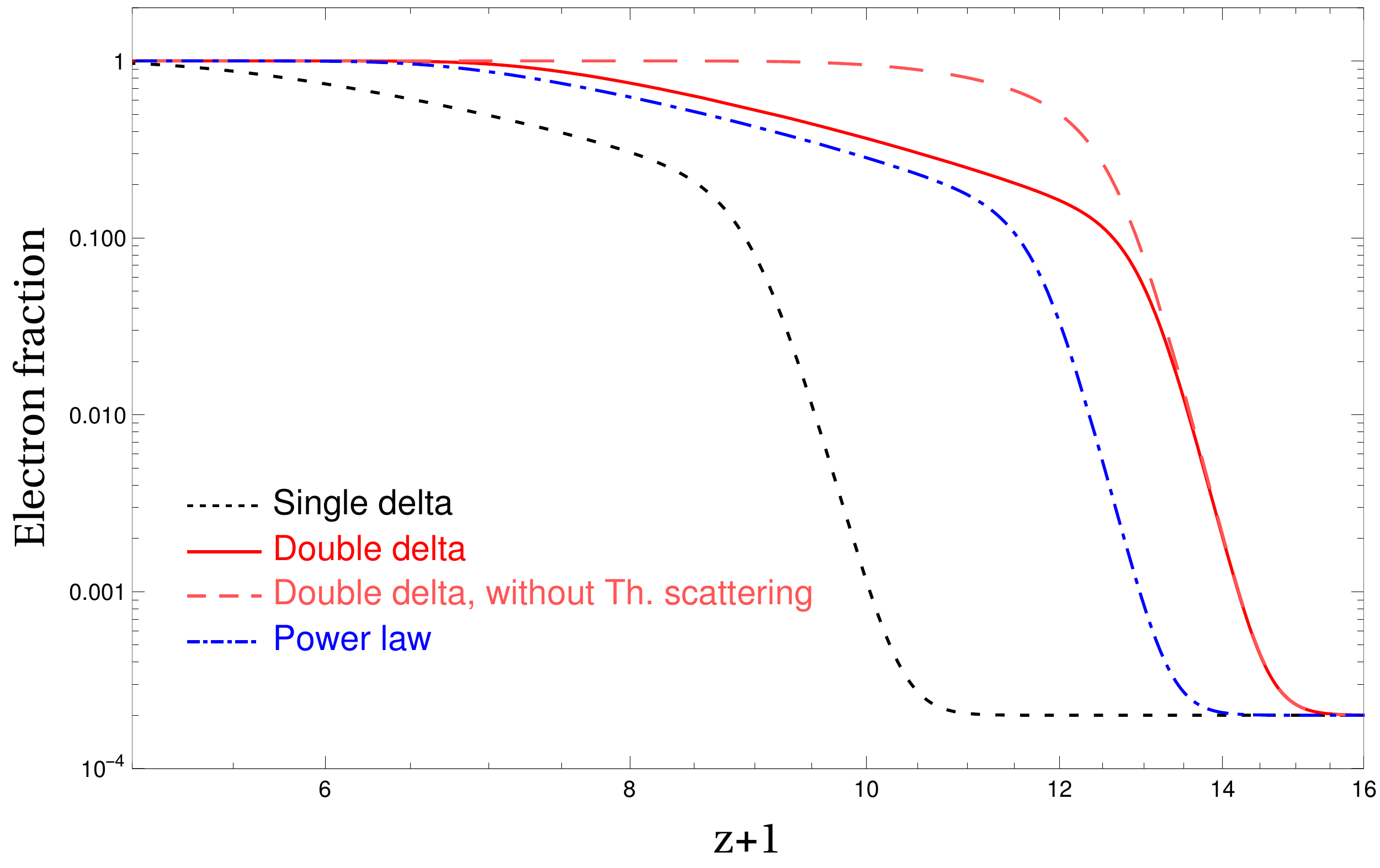}
			\caption{The temperature (left) and degree of ionization of matter (right)  depending on the redshift for the cases, when the  PBH density concentrated in one (black line) and two different mass values (red line) or distributed in mass as power-law (one of the best case). 
			} 
			\label{leftion}
		\end{center}
	\end{figure}

The temperature of baryonic matter and degree of its ionization for the PBH mass value ($M\approx 5 \times 10^{16}$ g), giving the best effect, are shown in figure~\ref{leftion} (left and right respectively) in dependence on the redshift.

The delta-functional mass spectrum approximates a continuum spectra containing one sharp maximum. Additional maximum in a mass spectrum needs another delta-function to be involved. Let us consider less trivial mass distributions.

In case of two peaks in mass spectrum, we took the first peak at the same position as in case of single delta-function mass spectrum (at $5 \times 10^{16}$ g) and added the second one at $7 \times 10^{16}$ g. The height of second peak was step-by-step raised, while simultaneously the first one was lowered in so manner that DGRB was saturated due to contributions from the both peaks. We stopped when reionization effect became maximal. In fact, taking first peak at any relevant mass and adding the second one leads to amplification of ionisation effect, so there is no big fine tuning here.
	The pair of mass values ($5 \times 10^{16}$ g and $7 \times 10^{16}$ g) provides one of the strongest amplification among those we looked over.
 
A width of peak could be extra parameter but we did not study it explicitly (though, an uniform mass distribution with a finite width is considered below as particular case) and can refer to \cite{1701.07223} on this issue.

Evolution of the temperature and degree of ionization of matter in the case of two peaks mentioned above is shown in the figure~\ref{leftion} together with other cases.

As one can see from figure~\ref{leftion} (left)  the temperature of baryonic matter $T$ begins to grow at the moment  $z\simeq 50$. At this time, the heating rate of matter due to ionization losses becomes higher than the rate of the Universe expansion. 
The dashed red line corresponds to the case when the interaction of baryonic matter (free electrons) with the CMB photons is neglected. It shows that interaction with the CMB becomes important when free electrons appear, what causes them to cool (at $z\sim10$).

Figure~\ref{leftion} (right) shows that for the considered double delta-function case, reionization can be reached by the moment $z\sim 6$, while for single delta function it was not earlier than $z\sim 4$. The given two delta functions provide contribution to the dark matter equal to 10\%. This contribution can be enhanced by a simple addition of a third peak beyond considered mass interval.
	
	\section*{Power-law mass distribution}

		 The mechanism of PBH formation due to collapse of domain walls, which are supposed to
		 form in the result of phase transitions at the inflationary stage \cite{1}, may give a variety of PBH mass distributions in dependence on initial parameters of a scalar field potential responsible for a phase transition. The form of PBH mass spectrum strongly depends on the form of potential. 
		 Nonetheless, simple forms of the latter lead, as a rule, to power-law-like form
		 \begin{equation}
		 \frac{dw}{dM}\propto M^{\alpha}
		 \end{equation}
		 with negative exponent about $-3\lesssim\alpha\lesssim-1$ (what is noted for other mechanisms in \cite{Carr}) and even with positive ones within limited most contributing mass range. But one should take into account a ``renormalization'' of mass distribution connected with possible successive coalescence evolution of PBH systems \cite{Dokuch}.
		 
		 Here we consider reionization effect from pure power-law mass spectra of PBHs with different typical exponents. 
		 Different mass intervals for each power-law distributions are considered ($M_\text{min}<M<M_{17}$, where $10^{15}~\text{g}<M_\text{min}<M_{17}$).
		 So we have two basic varying parameters: $\alpha$ and $M_{\min}$.
		 
	As earlier, gamma-ray flux is calculated for PBHs with given extended power-law mass distribution to constrain their maximal density from observation data. For the obtained maximal density, reionization effect is estimated. Evolution of ionization degree for one of the best cases (with highest ionization) of power-law mass distributions is shown in the figure~
	\ref{leftion} (right) in comparison with best delta-function-like cases.						
	
\begin{figure}
	\begin{center}
			{\includegraphics[scale=0.6]{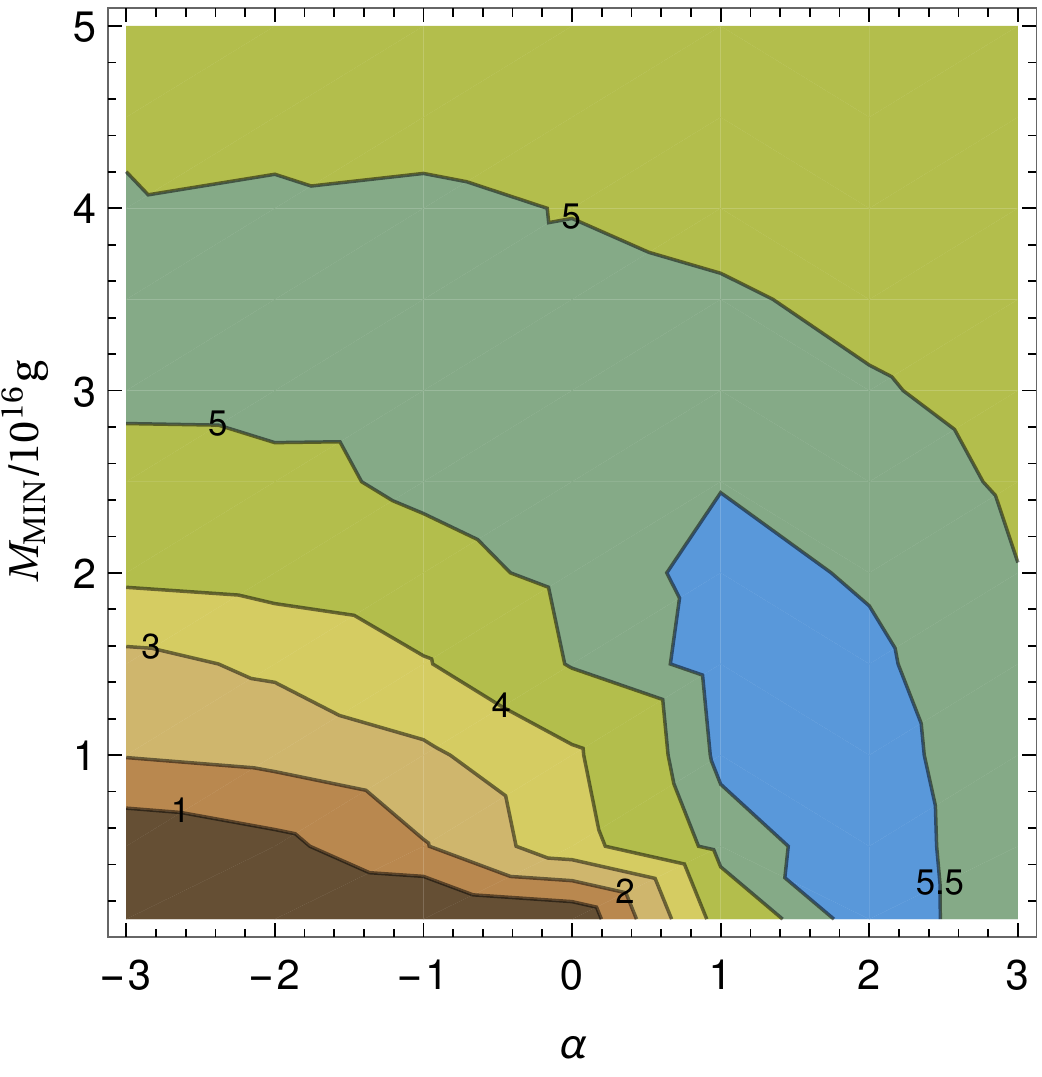}} 
			\qquad
			{\includegraphics[scale=0.6]{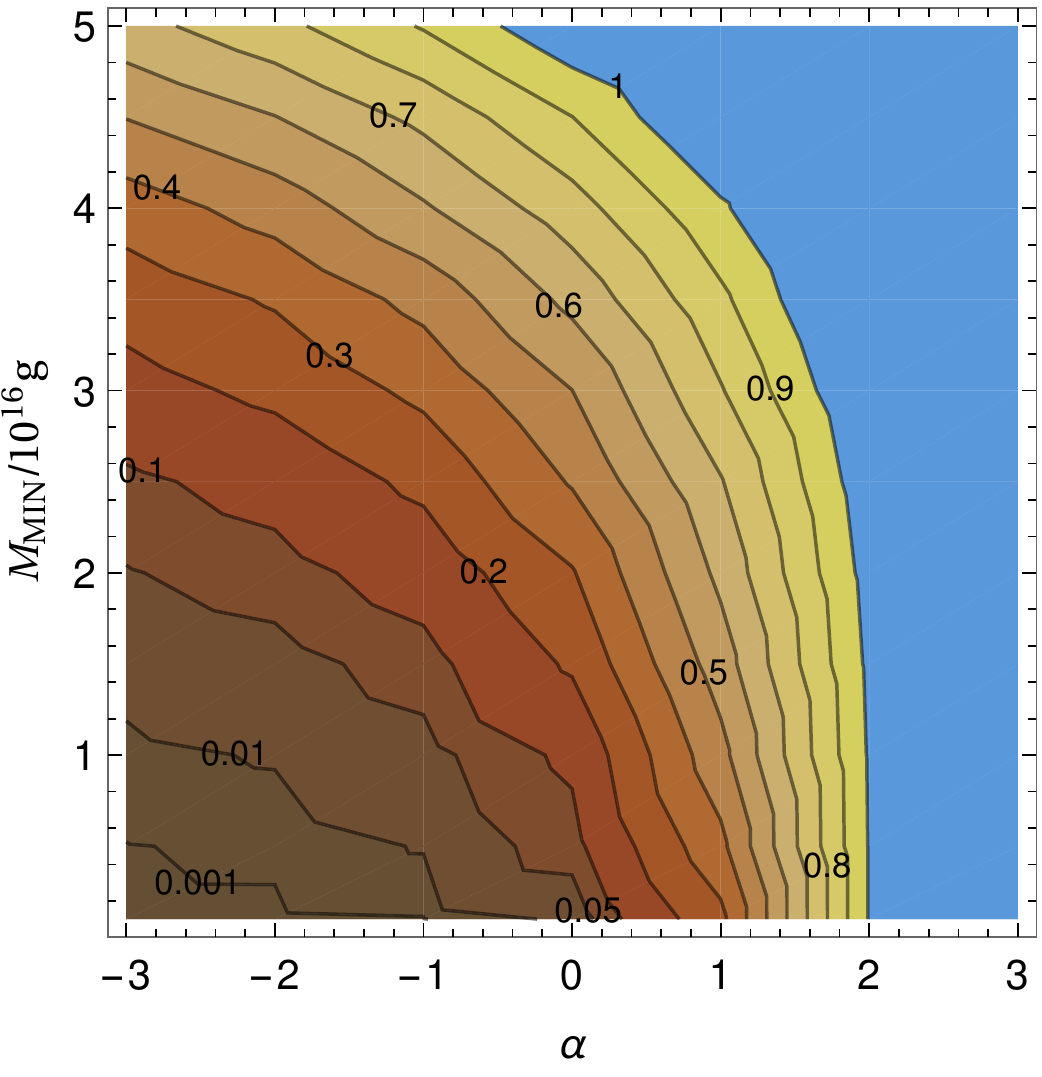}}
		\caption{The redshift (left), at which which 80\% of matter are ionized due to PBHs with power-law mass spectrum at $M_{\min}\lesssim M \lesssim M_{17}$ (for $\alpha>2$ see comments in the text), and their contribution to DM density (right) in dependence from $M_{\min}$ and $\alpha$.}
		\label{power}
	\end{center}
\end{figure}

	
	Changing $\alpha$ and $M_{\min}$ we get different ionization effect and contribution to DM.
	The figure~\ref{power} of left panel shows a redshift, at which 80\% of matter are ionized due to PBHs, for different values of $M_{\min}$ and $\alpha$. Right panel shows, for the same values, PBH contribution to DM density $\Omega_{\rm PBH}/\Omega_{\rm CDM}(M_{\min}<M<M_{17})$. 
	
	In the region $\alpha\sim 2\div 3$ (growing power-law spectra), full value $\Omega_{\rm CDM}=0.26$ can be reached, because high mass tail starts to contribute strongly. In this case PBH density distribution gets constrained due to the value $\Omega_{\rm CDM}=0.26$ rather than by DGRB. It means that PBHs do not saturate DGRB constraint thereby losing an ionization capability. To rescue such a loss, we change upper limit $M_{\max}=M_{17}$ in interval $2\lesssim\alpha\lesssim 3$ (exacter in all blue region of figure~\ref{power} (right)) by other value $M_{\max}<M_{17}$ to reach maximal PBH gamma-radiation while $\Omega_{\rm PBH}=\Omega_{\rm CDM}$. 
	It is obtained that $M_{\max}\approx (0.9-1)\times M_{17}$, and $z$, at which 80\% of matter ionized, increases by 1--2 as compared to the case $M_{\max}= M_{17}$. It possibly expands a gap in mass range before FC is on. Figure~\ref{power} includes these corrections.

	As one can see, there is a big region where reionization effect can be reached (better than in case of single-delta-function-like mass spectrum). 
	At the same time, PBH mass spectra with $\alpha\gtrsim 2$ may provide all DM density.
	
	One notes generally, we have taken a milder constraint from gamma-background (2--3 sigma excess over observational data was accepted in deriving restriction), and upto $M=M_{17}$ FC is ignored. The latter indulgence allowed to reach aforementioned possibility to get total DM density in form of PBHs within considered mass range. If one supposes FC takes a power at $M\approx 6\times 10^{16}$~g then DM 
	remains unexplained while reionization effect is weakened insignificantly (the left plot in figure~\ref{power} changes just a little). Nevertheless even in this case, one can try to get dark matter at the cost of some loss of ionization effect. If we look again at the left plot in figure~\ref{power} in the region for $M_{\min}\sim (1.5-2)\times 10^{16}$~g and $\alpha$ around $-2$, reionization is reached there at $z\sim 3-4$ which is not worst. But given power-law spectrum, each logarithmic mass interval ($\Delta \lg M=1$) gives approximately equal contribution to density as small as  $\Omega_{\rm PBH}/\Omega_{\rm CDM}\sim 0.01-0.1$ (see respective region in figure~\ref{power} (right)).
	So, if we extrapolate spectrum $M^{-2}$ for higher mass, then we get contribution $\Omega_{\rm PBH}/\Omega_{\rm CDM}\sim 0.03-0.1$ on each 
	interval $\Delta \lg M=1$ and hence reach the total DM density in about 10 intervals. 
	But it can have tension with FC, which covers $2.5-3.5$ orders of mass magnitudes and falls to  $\Omega_{\rm PBH}/\Omega_{\rm CDM}\sim 10 $\% in its minimum for delta-function-like mass distribution (so in the worst we should avoid contribution bigger than $10\%/(2.5-3.5)=3-4$\% in each $\Delta \lg M=1$ interval). However, we suppose that a special analysis of femtolensing effect for an extended PBH mass distribution is required to resolve situation (when all PBHs are assumed to have different mass values, it could be more difficult to reach statistically reliable result analysing data on gamma-ray bursters and constraint could be weakened). 
		
\section*{Conclusion}

	In our work we considered different forms of the PBH mass distributions. It was found 
	that the extended (not single delta-functional) PBH mass distribution gives greater contribution to the reionization of matter. In particular, the ``simplest'' complication of single delta-function mass spectrum by adding a second delta-function (adjusting simultaneously both peak heights to DGRB constraint) allows to enhance noticeably reionization effect. 
	A region of the values of $\alpha$ and $M_{\min}$ for power-law mass spectrum ($dw/dM\propto M^{\alpha}$, $M_\text{min}<M<M_{17}$) is found where reionization effect is maximal. Moreover, a part of that region with $\alpha>2$ can provide essential contribution to DM still evading femtolensing constraint (FC), if the latter is off until $M\lesssim M_{17}$. If FC takes a power at smaller $M$, then DM explanation possibility is mostly lost while reionization effect is weakened rather insignificantly over all considered $\alpha-M$ parameter space. But there is a benchmark region around $\alpha\sim-2$ of that space where DM explanation could be restored avoiding FC, what can be proved 
	by specific analysis of femtolensing effects for an extended PBH mass distribution.


\section*{Acknowledgment}
This work was supported by Russian Science Foundation and fulfilled in the framework of MEPhI Academic Excellence Project (contract \textnumero~02.a03.21.0005, 27.08.2013) and according to the Russian Government Program of Competitive Growth of Kazan Federal University.
The work of S.~G.~R. was also supported by the Ministry of Education and Science of the Russian Federation, Project \textnumero~3.4970.2014/BY.

\end{document}